\documentclass[journal]{IEEEtran}
\usepackage{amsmath,amssymb,amsfonts}
\usepackage{algorithmic}
\usepackage{graphicx}
\usepackage{textcomp}
\usepackage{caption}
\usepackage[utf8]{inputenc}
\usepackage[english]{babel}
\usepackage{subfig}
\usepackage{url}
\usepackage{booktabs}
\usepackage[ruled,vlined]{algorithm2e}
\usepackage[table,xcdraw]{xcolor}
 
\ifCLASSINFOpdf
\else
\fi
\begin{document}

\title{Designing constraint-based false data injection attacks against the unbalanced distribution smart grids}

\author{Nam~N.~Tran,
        Hemanshu~R.~Pota,
        Quang~N.~Tran,
        and~Jiankun~Hu$^*$,~\IEEEmembership{Senior Member,~IEEE}
\thanks{N. N. Tran, H. R. Pota, Q. N. Tran. and J. Hu ($^*$Corresponding Author) are with the School of Engineering and Information Technology, University of New South Wales Canberra at ADFA, ACT 2602, Australia (e-mail: nam.tran@student.adfa.edu.au; h.pota@adfa.edu.au; quang.tran@student.unsw.edu.au; j.hu@adfa.edu.au).}
\thanks{Copyright (c) 2021 IEEE. Personal use of this material is permitted. However, permission to use this material for any other purposes must be obtained from the IEEE by sending a request to pubs-permissions@ieee.org.}}

\markboth{ }%
{Shell \MakeLowercase{\textit{et al.}}: Bare Demo of IEEEtran.cls for IEEE Journals}

\maketitle

\begin{abstract}
The advent of smart power grid which plays a vital role in the upcoming smart city era is accompanied with the implementation of a monitoring tool, called state estimation. For the case of the unbalanced residential distribution grid, the state estimating operation which is conducted at a regional scale is considered as an application of the edge computing-based Internet of Things (IoT). While the outcome of the state estimation is important to the subsequent control activities, its accuracy heavily depends on the data integrity of the information collected from the scattered measurement devices. This fact exposes the vulnerability of the state estimation module under the effect of data-driven attacks. Among these, false data injection attack (FDI) is attracting much attention due to its capability to interfere with the normal operation of the network without being detected. This paper presents an attack design scheme based on a nonlinear physical-constraint model that is able to produce an FDI attack with theoretically stealthy characteristic. To demonstrate the effectiveness of the proposed design scheme, simulations with the IEEE 13-node test feeder and the WSCC 9-bus system are conducted. The experimental results indicate that not only the false positive rate of the bad data detection mechanism is 100 per cent but the physical consequence of the attack is severe. These results pose a serious challenge for operators in maintaining the integrity of measurement data.	
\end{abstract}


\IEEEpeerreviewmaketitle

\section{Introduction}\label{introduction}

\IEEEPARstart{I}{n} the era of smart cities, the intelligent systems of various Internet-of-Things based applications such as smart grid, smart transporation, and smart health-care are expected to interconnect with each other \cite{lin2017survey}. Undertaking the task of monitoring and control in the smart power grid is the energy management system (EMS) whose state estimation (SE) is responsible for collecting and processing measured data. A set of sensors and measurement devices are placed at appropriate positions to acquire both analog and digital data, serving as the inputs for estimating the state variables. Based on the set of estimated state variables, a snapshot of the current status of the grid is obtained, providing insight for making control decisions. Since the measured data are possibly contaminated by noise, a bad data filtering mechanism is often equipped to eliminate grossly erroneous data.

In the past, SE module was only implemented for the fully balanced interconnected transmission system where the data are collected and transmitted to a centralized control center. With the increasing penetration of distributed energy resources (DERs), the implementation of SE module for the unbalanced residential distribution grid is indispensable. At this level, however, the SE module only collects and processes measurement data from a specific part (for instance, several feeders from a substation) rather than the whole grid, i.e., operating in a local manner. Given this distinctive attribute, applying the edge computing is more efficient than the cloud computing as the enormous amount of data do not need to be transmitted to the centralized computer system to process, enhancing the module performance. This feature is critically important as an SE module for distributed grid must always employ a full nonlinear AC analysis instead of linearized DC-based measurement model.

Since the SE module relies on the cyber-domain to collect measurement data from distances, the resultant outputs can be manipulated. As a major IoT system, the Ukraine 2015 incident is a typical IoT network attack \cite{yu2018security}. Recently, a new type of cyber-attack targeting SE module has been discovered, which is called false data injection (FDI) attack. Liu \textit{et al.} \cite{liu2011false} demonstrated that a well-designed data-driven FDI attack is capable of bypassing the bad data filtering mechanism of the SE module, potentially producing harmful consequences to the physical domain of the smart grid. This paper focuses on the realization of the FDI attack targeting the unbalanced residential distribution grid. An attack design scheme which is based on physical constraint to exhibit the theoretically stealthy characteristic is presented. The FDI design is applied to a case study with the IEEE 13-bus test feeder. In addition, the consequence of the FDI attack are also discussed, indicating how severe the system suffers under the impact of the attack. A comprehensive insight about this FDI attack enables various strategies to mitigate its effect or its prevention can be planned. There are seven sections in this paper. A thorough review regarding the related work is presented in Section \ref{rw}. Section \ref{FDI} presents FDI attack design principles that must be considered while working with distribution grid and steps in planning the generalized design schemes for specific cases. In addition, a method for assessing the SE's output which is based on the unbalanced load flow program is also devised to quickly evaluate the stealthy characteristic of the attack data. Section \ref{case} provides a case study with the unbalanced IEEE 13-node test feeder to illustrate the proposed attack design schemes. The experimental results are presented in Section \ref{exre}, demonstrating the theoretical stealthy capability of the attack. Another case study with the WSCC 9-bus system is presented to provide a glimpse about the impact that an FDI cyber-attack imposes on the physical environment. Finally, concluding remarks and future research trends are highlighted in Section \ref{conclu}. 
\section{Related work}
\label{rw}
The development of the SE module for unbalanced distribution grid first concentrated on improving performance of the optimization algorithms and efficiency of the bad data filtering mechanism. Since the first released prototype in 1970 \cite{schweppe1970power1}\cite{schweppe1970power2}\cite{schweppe1970power3}, several research groups proposed various paradigms for the SE module operating at the medium-and-low-voltage distribution grid. Roytelman \textit{et al.} \cite{roytelman1993state}, Baran \textit{et al.} \cite{baran1994state}, Meliopoulos \textit{et al.} \cite{meliopoulos1996multiphase} and Lu \textit{et al.} \cite{lu1995distribution} obtained extensive full-phase power based models. After that, Baran \textit{et al.} improved the computing efficiency of the module by introducing a current-based model \cite{baran1995branch}. Using the same ideas about the current-based model, Wang \textit{et al.} \cite{wang2004revised} also enhanced the working rate of the SE module. The impact of noise on the performance of the SE module was also carefully studied. Most notably, Monticelli \textit{et al.} \cite{monticelli1983reliable}, Van Cutsem \textit{et al.} \cite{van1984hypothesis}, Singh \textit{et al.} \cite{singh2005topology}, and Gou \textit{et al.} \cite{gou2014unified} proposed various techniques to eliminate bad data, including hypothesis testing and fuzzy pattern matching. However, theses algorithms were planned to deal with sets of measurements with independent errors only. Since the event of Stuxnet \cite{stuxnet} in 2010 and the explore of false data injection attack by Liu \textit{et al.} \cite{liu2011false} in 2011, the man-in-the-middle data-driven attack has been getting more and more attentions. This method of systematically manipulating data has proved its effectiveness in bypassing the traditional bad data detector.

The methods to coordinate the attack data are studied in various publications. Yang \textit{et al.} \cite{yang2013false}, Anwar \textit{et al.} \cite{anwar2017modeling}, Yu \textit{et al.} \cite{yu2015blind}, Rahman \textit{et al.} \cite{rahman2012false}, and Liu \textit{et al.} \cite{liu2015modeling} investigated attack cases with realistic resource-limited constraints to demonstrate the feasibility of FDI attack in different situations. While most of research groups worked on the linearized DC-based measurement model, the others such as Hug \textit{et al.} \cite{hug2012vulnerability}, Jia \textit{et al.} \cite{jia2012nonlinearity} and Rahman \textit{et al.} \cite{rahman2013false} necessitated the deployment of the nonlinear AC-based model in their corresponding researches. To this extend, the FDI attack was considered as a serious cyber-threat to all level of power grid, regardless of transmission or distribution system \cite{deng2018false}. As such, the topic of FDI attack agaist distribution grid has gained as much interest as at the transmission level.

Since the SE module for the distribution grid operates locally based on the edge computing with scattered measurement devices that are more approachable in terms of physical placement, these devices are likely to be compromised. Sharing this same vision about cyber-threat in cyber-physical system, Lim \textit{et al.} \cite{lim2009security} proposed a set of security protocols to deal with cyber-attacks for the distribution system. Meanwhile, Guo \textit{et al.} \cite{guo2013online}, Liu \textit{et al.} \cite{liu2016microgrid}, and Berg \textit{et al.} \cite{beg2017detection} provided security analyses related to the impact and strategies to mitigate the cyber-attack consequences in the presence of the DERs. Recently, the application of moving target defense (MTD) in designing FDI attack detection solution has achieved some positive results through the effort of Lin \textit{et al.} \cite{lin2018raincoat} and Jhala \cite{jhala2020perturbation}. Taking deeper analysis on attack techniques, Yang \textit{et al.} \cite{yang2013novel}, Lin \textit{et al.} \cite{lin2012false}, Liu \textit{et al.} \cite{liu2017false}, and Aoufi \textit{et al.} \cite{aoufi2020survey} assessed various scenarios of false data injection attack disturbing the normal operation of the distribution grid. Deng \textit{et al.} \cite{deng2018false} chose to focus on the technique aspect of FDI attack in terms of data which is based on the nonlinear balanced measurement model. From that foundation, Zhuang \textit{et al.} \cite{zhuang2019false} developed an attack procedure for multiphase unbalanced system whose measurement model is linearized. In \cite{cpe.5956}, an AC model FDI attack is successfully designed for the interconnected transmission system. Inspired by this work, in this paper it is extended to the unbalanced residential network. A comprehensive attack design scheme against the nonlinear unbalanced SE module as well as a preliminary investigation of the consequences of the attack as in this paper fulfill gaps in the literature.

\section{False Data Injection Attack Design Scheme for Distribution Grid}
\label{FDI}
\subsection{Design considerations}
\label{conside}
In order to actively formulate strategies to deal with the potential threats of the FDI attack, the worst-case scenario must be taken into consideration. A typical example for this situation is the 2015 Ukraine blackout \cite{liang20162015}\cite{case2016analysis}. That incident consists of a series of well-prepared destructive actions, from phishing email to seize the control of SCADA (cyber domain) to remotely issuing the commands to switch off the power substations (physical domain). In this research, we assume that the adversary successfully hijacks the SCADA system, conducts reconnaissance, and gets access to the firmware of the RTUs. As a result, the adversary is capable of identifying and collecting necessary information about the attack region (a sub-grid of the network) including topology, status of switches and breakers, steady-state measured values, and also to compromise whichever critical reading that is needed to be overridden. We also assume that the measurement devices are placed at suitable positions not only to guarantee the system observability but also to improve the performance of the SE module \cite{singh2009measurement}. The main task of the attacker is then to construct a completed set of false data that look like normal measured data from the viewpoint of the SE module. These are the standard assumptions made in most of the existing literatures. In this paper, the focus is on technical aspects of computing FDI attack vectors with the ability to completely bypass the bad data detector (BDD) of the SE module.
\subsubsection{Models of components}
Distribution grid has no phase-transposing, is mostly radial in structure with the number of phases on each lateral varies from one to three; and most importantly, it has naturally unbalanced load. Both the representative models and analytic algorithms applied for distribution level are much different compared to the transmission system. For instance, as the phase transposing is not applied, the combination of “self“ and “mutual induction” into “phase induction” is no longer valid which does not reduce to decoupled single-line system as for transmission systems. Instead, the computation on each quantity must be conducted independently using Carson’s equations \cite{kersting2006distribution} that result in individual self and mutual impedances (\textbf{\textit{of simplified model}}) presented in the matrix format as below:
\begin{equation}
	Z_{abc} = 
	\begin{bmatrix}
		Z_{aa}&Z_{ab}&Z_{ac}\\
		Z_{ba}&Z_{bb}&Z_{bc}\\
		Z_{ca}&Z_{cb}&Z_{cc}\\
	\end{bmatrix}\
\end{equation}
\subsubsection{Models of measurements}
Given the fact that the distribution smart grid is revolutionized with high penetration of DERs, it now has bidirectional power flows and a meshed structure \cite{cruz2018meshed} (but sometimes operates in radial mode). Therefore, it is reasonable to expect that the SE module for the distribution grid has the same properties and characteristics as of the transmission system. Let's assume that the RTUs can collect measurement quantities (active and reactive powers \cite{milanovic2013international}\cite{baran2009distribution}\cite{thornley2005state}) at both the terminals of each phase on branches and at positions where power is injected so the redundancy level is high enough to guarantee the system observability. In the past, limited number of RTUs and the low quality of measurements were the major obstacle in implementing the state estimation for distribution grid. These issues are solved by applying recent techniques, for instance, augmenting matrix completion with power flow constraints, incorporating heterogeneous data sources \cite{donti2019matrix}, and employing the historical measurements \cite{abdel2012low} to compensate for the paucity of high-quality measurement data. Given the unbalanced characteristics of the distribution network, employing per-phase representative model in designing FDI attack is inadequate. Instead, a full 3-phase power-flow measurement model as in Eq. (\ref{3m1}) and (\ref{3m2}) that includes all the unique features of the distribution network, is employed: 
\begin{equation}
	\begin{split}
		P^{ph}_{ij} & = \sum_{l = a,b,c} V^{ph}_i\{V^l_i[G^{ph,l}_{ij}\cos(\theta ^{ph}_i - \theta ^l_i) + \\
		& B^{ph,l}_{ij}\sin(\theta ^{ph}_i - \theta ^l_i)] \} \\
		& - \sum_{l = a,b,c} V^{ph}_i\{V^l_j[G^{ph,l}_{ij}\cos(\theta ^{ph}_i - \theta ^l_j) + \\ & B^{ph,l}_{ij}\sin(\theta ^{ph}_i - \theta ^l_j)] \}
	\end{split} 
	\label{3m1}	
\end{equation}
\begin{equation}
	\begin{split}
		Q^{ph}_{ij} & = \sum_{l = a,b,c} V^{ph}_i\{V^l_i[G^{ph,l}_{ij}\sin(\theta ^{ph}_i - \theta ^l_i) - \\
		& B^{ph,l}_{ij}\cos(\theta ^{ph}_i - \theta ^l_i)] \}\\
		& - \sum_{l = a,b,c} V^{ph}_i\{V^l_j[G^{ph,l}_{ij}\sin(\theta ^{ph}_i - \theta ^l_j) - \\ & B^{ph,l}_{ij}\cos(\theta ^{ph}_i - \theta ^l_j)] \}
	\end{split} 
	\label{3m2}	
\end{equation}
where \textit{i}, \textit{j} = 1, 2, ..., N with N is the total number of nodes, and \textit{ph} denotes phase \textit{a}, \textit{b}, or \textit{c}. \textit{G} and \textit{B} are the real and imaginary parts of the elements of $Z_{abc}$.

All the three phases in the distribution grid are coupled, which means that any change happens in one phase results in alterations in the quantities of the other phases. This feature makes the attack design scheme for distribution grid more challenging than the counterpart in transmission system since the latter one does not have to consider that interaction.
\subsubsection{Models for the State Estimation}
The relationship between measurements and state variables is given as:
\begin{equation}
	z = h(x) + \epsilon
	\label{state}
\end{equation}
where
\begin{itemize}
	\item \textit{z} is a vector of measurements (power flows, power injections, voltage magnitudes and angles). In distribution grid, all the measurements are phase quantities.
	\item \textit{x} is a vector of state variables (voltage magnitudes and angles at every phase of each node).
	\item \textit{h} is a vector of nonlinear functions representing the relationship between measurement values and state variables.
	\item $\epsilon$ is a vector of measurement errors which is assumed to have Gaussian distribution with zero mean.
\end{itemize}
Given a set of measurements, an SE algorithm iteratively solves, for instance, the most common weighted least square optimization problem as in (\ref{eq3.3}), to obtain the set of state variables $ \bar{x} $ \cite{monticelli2012state}:
\begin{equation}
	\label{eq3.3}
	min\mathbf{F}(\bar{x}) = (z - h(\bar{x}))^T\cdot\mathbf{W}\cdot(z - h(\bar{x}))
\end{equation}
In the above equation, \textbf{W} is the weighting matrix whose elements correspond to the inverse of the individual measurement accuracy. The existence of bad measurement data due to various reasons is detected once the normalized residual exceeds a threshold $\tau$ (which is defined based on the degree of fault tolerance), i.e.,
\begin{equation}
	\label{eqtau}
	||z - h(x)|| > \tau
\end{equation}
\subsection{Design scheme}
Corresponding to linear DC-based and nonlinear AC-based SE models, the attack models are also divided into DC-based and AC-based attacks. The DC-based FDI attack model can easily bypass the criterion (\ref{eqtau}) just by adding an attack vector \textit{a} that is the product of the linearized matrix \textbf{H} and an arbitrary contaminated vector \textit{c} to the current measurement vector \textit{z}. However, the FDI attack aimed at an AC-based SE module is far more complicated. This type of attack targets a specific region, called attack area. Let the whole grid be divided into two regions: region 1 is the area under attack and the rest of the network is region 2. The corresponding sets of measurements, state variables and nonlinear representative functions are \{$ z_1, z_2 $\}, \{$ x_1, x_2 $\}, and \{$ h_1, h_2 $\}, respectively. The relationship between measurements and state variables becomes:
\begin{equation}
	\begin{bmatrix}
		z_1\\
		z_2
	\end{bmatrix}
		=
	\begin{bmatrix}
		h_1(x_1, x_2)\\
		h_2(x_2)
	\end{bmatrix}
		+
	\begin{bmatrix}
		\epsilon _1\\
		\epsilon _2
	\end{bmatrix}
\end{equation}

The vector \textit{h}(\textit{x}) is based on physical laws such as the Kirchhoff’s current law and the Kirchhoff’s voltage law (both originate from the law of conservation of energy), and Ohm’s law. Any set of values $\hat{z}$ which strictly complies with the physical laws that constitute the measurement models, is able to provide a solution for the optimization problem (\ref{eq3.3}). If another set of this kind exists in the proximity of a steady state operating point, that set of measurements is also able to bypass the examining process of the BDD.

In this paper, the above FDI attack design principle is applied, in which the set of manipulated state variables $\hat{x}$ = \{$\hat{x_1}, x_2 $\} and the set of malicious measurements $\hat{z}$ = \{$\hat{z_1}, z_2 $\} will be obtained directly. A new pseudo steady state near the genuine one is created to deceive the SE module. Because it is a pseudo steady state, it meets the criterion (\ref{eqtau}):
\begin{equation}
	\Bigg\|
	\begin{bmatrix}
		\hat{z_1}\\
		z_2
	\end{bmatrix}
	-
	\begin{bmatrix}
		h_1(\hat{x_1}, x_2)\\
		h_2(x_2)
	\end{bmatrix}
	\Bigg\| < \tau 
\end{equation}

The above proposed method is fundamentally different from all the previous approaches (for balanced transmission system) in the literature \cite{hug2012vulnerability}\cite{jia2012nonlinearity}\cite{rahman2013false} as those algorithms try to find the contaminated vector \textit{c}, the set of manipulated state variables $ \hat{x} = \{x_1 + c, x_2\} $, and the set of mixture of malicious and normal measurements $ \hat{z} = \{z_1 + a, z_2\} $, respectively in that order. The normalized residual is then qualified if:
\begin{equation}
	\Bigg\|
	\begin{bmatrix}
		z_1 + a\\
		z_2
	\end{bmatrix}
	-
	\begin{bmatrix}
		h_1(x_1 + c, x_2)\\
		h_2(x_2)
	\end{bmatrix}
	\Bigg\| < \tau 
\end{equation}
Hence, the condition for the attack being hidden is:
\begin{equation}
	\begin{split}
		&a - h_1(x_1 + c, x_2) + h_1(x_1, x_2) = 0\\
		&\Leftrightarrow a = h_1(x_1 + c, x_2) - h_1(x_1, x_2)
	\end{split}
\end{equation}
Attack vector \textit{a} and contaminated vector \textit{c} are mutually dependent for a successful attack. Therefore, all elements of the contaminated vector \textit{c} cannot be arbitrarily selected as Anwar \textit{et al.} claimed in \cite{anwar2014vulnerabilities}. In addition, the approach in \cite{anwar2014vulnerabilities} could possibly produce cumulative errors due to enduring through various computation stages.

\begin{figure}[h]
	\centering
	\includegraphics[scale=0.75]{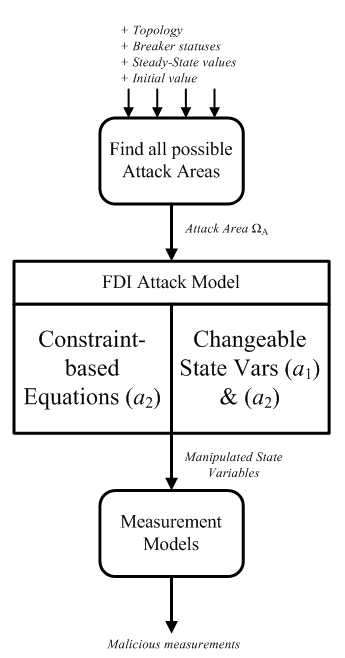}
	\caption{The FDI attack design scheme for distribution network.}
	\label{fig0}
\end{figure}

The proposed attack design scheme in this paper, shown in Fig. \ref{fig0}, follows the same two-stage process/principle as in \cite{cpe.5956} that was designed for the transmission system. For the unbalanced distribution grid, the design scheme is distinguished by: (\textit{i}) taking the unbalanced characteristics of the grid into consideration; and (\textit{ii}) devising appropriate algorithms to identifying the attack components such as the coverage of impact region and the attack model (the bus admittance matrix for a whole distribution grid is not available for the algorithm to define attack areas, and the appearance of center-tapped transformer introduces additional constraint equations to the attack model). The first stage is to identify all possible attack areas. Within an attack area, the measurements are compromised. The first condition to guarantee a successful attack is that the attack region must be enclosed by nodes with power injection only \cite{hug2012vulnerability}. This is important as the type of attack area helps to allocate resources for launching attack. In the next step the constraint-based attack model is obtained for the attack areas from the attack area identification stage. The output of this step is a set of manipulated state variables of a false steady state. This set is fed into the measurement model to compute the set of measurements with false data injection.
\subsubsection{Attack Area}
The attack area should be as small as possible because of two reasons: (\textit{i}) a smaller attack area would require less resource to launch the attack, e.g. the number of readings that the adversary have to alter; (\textit{ii}) changes in a smaller area probably may not attract an operator attention, thus reducing the chance of being detected.

As discussed above, the necessary condition for an attack area to be invisible is to have a boundary of nodes with power injection only. Since an FDI attack will modify the state variables at various nodes, the involved measurements will be altered accordingly. All these alterations must be "assigned" to some nodes in order to avoid any inconsistency. Nodes with no power injection can only adjust by the changes in the power flow along the connected branch, thus making the attack area larger. Meanwhile, nodes with power injection (either incoming or outgoing) can justify any change without further expanding the attack area by adjusting the metered value of the local power injection. Algorithm \ref{Algo1} presents the entire process of identifying the attack area in a distribution network.
\begin{algorithm}
	\caption{Finding FDI Attack Area for Distribution Network}
	\label{Algo1}
	\KwIn{$Z^{ij}_{abc}$ for all branches \textit{ij}, Initial node \textit{a}}
	\KwOut{Area of Attack $\Omega_A$}
	\begin{algorithmic}
		\STATE \textbf{1} Scanning all $Z^{ij}_{abc}$
		\IF{($Z^{ka}_{aa} \neq 0$) or ($Z^{ka}_{bb} \neq 0$) or ($Z^{ka}_{cc} \neq 0$)}
		\STATE Add node \textit{k} to $\Omega_A$
		\ELSE \STATE \textit{k}++
		\ENDIF
		\STATE \textbf{2} Scanning $\Omega_A$:
		\IF{(typeofnode(\textit{i}) = injection)}
		\STATE Move to the next node in $\Omega_A$
		\ELSE \STATE Back to \textbf{1}
		\ENDIF
	\end{algorithmic}
\end{algorithm}

\begin{figure}[h]
	\centering
	\subfloat[Attack area does not contain no-injection node.]{
		\label{subfig:a1}
		\includegraphics[scale=0.65]{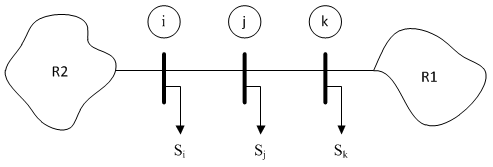} } \\
	\subfloat[Attack area contains node j as a no-injection node.]{
		\label{subfig:a2}
		\includegraphics[scale=0.65]{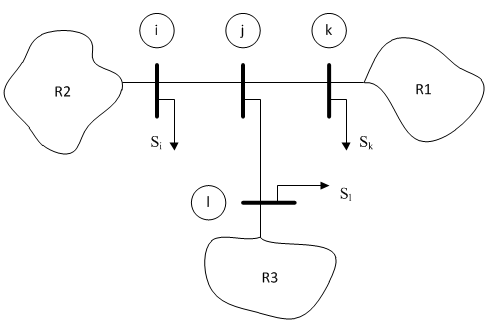} } 
	\caption{Two types of attack area exist in distribution network}
	\label{fig2}
\end{figure}

There are two types of attack areas, called ($a_1$) and ($a_2$) that exist within a distribution grid. The ($a_1$) attack area does not contain any no-injection node, and the ($a_2$) attack area contains at least one no-injection node. Fig. \ref{fig2} illustrates the general prototypes of these two attack areas. Although the only feature that separates two types of attack area is the existence of a node with power injection, the attack design schemes are quite different. The node with no power injection aggravates the complexity of the design problem as various physical-law based constraint equations must be formulated and solved. On the other hand, the design process for an attack area without no power injection node is more straightforward as it does not require the solution of the set of nonlinear equations (\ref{constr}), hence significantly reducing computations.
\subsubsection{FDI Attack Model}
\label{selsv}
From the set of nodes inside an attack area $\Omega _A$, we can calculate the number of changeable state variables. For every node in each phase, there are two accompanied state variables, voltage magnitude $V^{ph}_i$ and voltage angle $\theta^{ph}_i$. However, all the nodes on the boundary of the attack area must be removed from the set. If these nodes are also modified, the measurements on the connected branches, outside the attack area, will be altered as well. This might introduce inconsistencies that are likely detectable by the BDD. Furthermore, there is a type of state variable with predefined value that any imposed adjustment will attract attention, for instance, voltage magnitude and voltage angle of a slack bus. For that reason, the attack should always avoid those types of nodes. The complete set of all changeable state variable can be obtained through Algorithm \ref{Algo2}.
\begin{algorithm}
	\caption{Identifying changeable state variables}
	\label{Algo2}
	\KwIn{Area of Attack $\Omega_A$}
	\KwOut{The set of changeable state variable \textbf{SV}}
	\begin{algorithmic}
		\STATE No. of SV $n \leftarrow 2\times Size of(\Omega_A)$
		\STATE \textbf{SV} = \{$|V|_j, \theta _j| j \in \Omega_A$\}
		\WHILE{($j \in \Omega_A$)}
		\IF{(type of node (\textit{j}) = Slack)}
		\STATE \textit{n} = \textit{n} - 2
		\STATE Remove $|V|_j$ and $\theta _j$
		\ELSIF {(type of node (\textit{j}) = PV)}
		\STATE \textit{n} = \textit{n} - 1
		\STATE Remove $|V|_j$
		\ELSE
		\STATE \textit{j}++
		\ENDIF
		\ENDWHILE 
	\end{algorithmic}
\end{algorithm}

If the attack area is ($a_1$), the set of changeable state variable are solely located at the initial node. Given the omission of no power injection node, the attack design here goes straight to the stage of computing the set of false data. A new pseudo steady state is created, then all the new measurements are calculated accordingly. Any change on the branches are adjusted by means of changing the measured value of power injections at the local as well as the adjacent nodes.

Dealing with ($a_2$)-type attack area is a bit more complicated as it requires to apply the nonlinear constraint-based attack model to find the set of manipulated state variables. The model is represented by a set of constraint equations that has the basis in the law of conservation of energy. In order to launch an ideal stealthy FDI attack on a predefined attack area $\Omega_A$, all possible alterations must happen locally. It means that all power exchanges with the outside regions must be kept unchanged in order to maintain the seamlessly transition between the attack area and the rest of the grid. In addition, the characteristic of no power injection node must also be guaranteed. These requirements are satisfied by following the two rules below:
\begin{itemize}
	\item Algebraic sum of all power flows at a no power injection node must be equal to zero. By complying with this condition, the relationships that constitutes the \textit{h} function in (\ref{state}) are preserved.
	\item Regarding the attack area, the sum of all changes in every branches plus the sum of all injection variations must be equal to zero. By complying with this condition, the consistencies between powers inside and outside the attack area are preserved.
\end{itemize}
Given the set of no power injection node is $\Omega _{A0}$, the constraint-based FDI attack model is constructed as:
\begin{equation}
	\begin{aligned}
	&\begin{split}
		&\sum_{j \in \Omega _{A0}} \hat{z}^{ph}_{jk}(\hat{x_1}, x_2) = 0	\\
		&\sum_{j \in \Omega _{A0}} \Delta \hat{z}^{ph}_j(\hat{x_1}, x_2) + \sum_{i,j \in \Omega _A} \Delta \hat{z}^{ph}_{Lij}(\hat{x_1}, x_2) = 0 \\  
	\end{split}\\
	\textrm{s.t.} \quad & x_1^{i,min} \leq \hat{x_1^i} \leq x_1^{i,max}\\
	\end{aligned}
	\label{constr}
\end{equation}
where $x_1^{i,min}$ and $x_1^{i,max}$ are the minimum and maximum allowable values for state variable $i^{th}$ that indicates the steady state operation of grid.

As per-phase analysis is conducted, each rule must be composed for every phase in the attack coverage. In the same manner, a new pseudo steady state is created by an initialization, then this attack model outputs the manipulated state variables to compute the malicious measurements. By keeping the state variables at boundary unchanged, all the alterations are held inside the attack area. The changes of measurements on branches in the attack region ($P_{ij}$, $Q_{ij}$) will be justified by changing the measured values of the power injections at all power injection nodes. The active and reactive power flows are calculated by Eqs. (\ref{3m1}) and (\ref{3m2}), for all the phases in order to update the new meters’ values.

The explanations of related quantities, superscripts and subscripts using in the above constraint-based attack model are provided below:
\begin{itemize}
	\item $P^{ph}_{ijo}$, $Q^{ph}_{ijo}$ - Real and reactive power measured on phase \textit{ph} on branch from node \textit{i} to node \textit{j} that we obtained from load flow result (denoted by “\textit{o}”).
	\item $P^{ph}_{ijn}$, $Q^{ph}_{ijn}$ - Real and reactive power measured on phase \textit{ph} on branch from node \textit{i} to node \textit{j} that has been changed due to new state variables applied at one or both terminals of the branch (denoted by “\textit{n}”).
	\item $P^{ph}_{io}$, $Q^{ph}_{io}$ - Real and reactive power injection measured on phase \textit{ph} at node \textit{i} that we obtained from load flow result (denoted by “\textit{o}”).
	\item $P^{ph}_{in}$, $Q^{ph}_{in}$ - Real and reactive power injection measured on phase \textit{ph} at node \textit{i} that has been changed due to new state variables applied at terminal (denoted by “\textit{n}”).
	\item $P^{ph}_{Lijo}$, $Q^{ph}_{Lijo}$ - Power flow losses of phase \textit{ph} on branch \textit{ij} that we obtained from load flow result, $(P^{ph}_{Lijo} = P^{ph}_{ijo} + P^{ph}_{jio}, Q^{ph}_{Lijo} = Q^{ph}_{ijo} + Q^{ph}_{jio})$.
	\item $P^{ph}_{Lijn}$, $Q^{ph}_{Lijn}$ - Power flow losses of phase \textit{ph} on branch \textit{ij} that has been changed due to new state variables applied at one or both terminals of the branch, $(P^{ph}_{Lijn} = P^{ph}_{ijn} + P^{ph}_{jin}, Q^{ph}_{Lijn} = Q^{ph}_{ijn} + Q^{ph}_{jin})$.
\end{itemize}
%

\subsection{Assessment method}
\label{AMethod}
Given the increasing penetration of DERs that leads to a rising demand for real-time system monitoring of the distributed smart grid, the SE module is now indispensable. Although the adoption of SE module for unbalanced distribution grid is currently at a much lower level compared with the corresponding package for transmission system, but various prestigious power-oriented software vendors have recently developed dedicated module to handle the unbalanced state estimator \cite{magnago2016multiphase}\cite{krishnan2020validation}\cite{aoufi2020survey}\cite{dehghanpour2018survey}. However, for the purpose of evaluating the result of an attack design scheme, these expensive packages are not directly applicable. Instead a dedicated assessment process is devised based on empirical observation. The foundation of this assessment method is the consistency between the SE results and the load flow results. The FDI attack model generates a set of measurements that, in turn, will create a pseudo steady state that the state estimation routine considers as a genuine operating point. Thus, an FDI attack will completely bypass the bad data detection process if its loading values can generate the load flow results that match perfectly with the falsified values. A simulation tool from DIgSILENT, PowerFactory \cite{PF2017}, which is well-known for its capability of providing a comprehensive unbalanced load flow result, is selected to judge the designed attack. Fig. \ref{fig7} illustrates the general idea of the detection process. The details are sequentially presented below. Subscript 1 is used for manipulated values (quantities within the attack area) and subsript 2 for unchanged values (quantities outside the attack area) as in Section \ref{conside}.
\begin{figure}[h]
	\centering
	\includegraphics[scale=0.70]{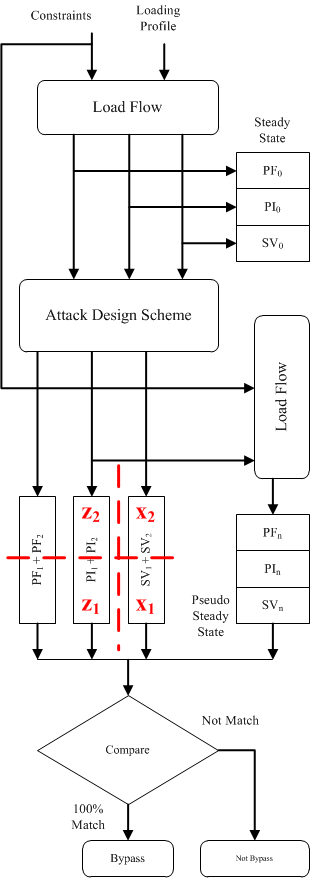}
	\caption{The flow chart of the assessment method that works similar to an unbalanced DNSE.}
	\label{fig7}
\end{figure} 
\subsubsection{Acquiring steady-state values from load flow, denoted by script 0}
These include steady-state measurements $z_0 = \{\mathbf{PI_0}, \mathbf{PF_0}\}$, and steady-state variables $x_0 = \{\mathbf{SV_0}\}$, which will be used as the input to the attack design process. \textbf{PI} is the power injection at nodes, either injected into (generated) or drawn from (consumed) a node. \textbf{PF} is the power flow measurement ($P_{ij}$ or $Q_{ij}$), indicating the algebraic power flows at the two terminals of each branch in the network (lines, transformers etc.). \textbf{SV} is the state variable (voltage magnitude or voltage angle).
\subsubsection{False data generation}
Feeding the steady-state sets $\{z_0, x_0\}$ into the attack design scheme, we obtained attack design results:
\begin{equation}
\begin{split}
&\hat{z} = z_1 + z_2 = (\mathbf{PF_1} + \mathbf{PF_2}) + (\mathbf{PI_1} + \mathbf{PI_2})\\
&\hat{x} = x_1 + x_2 = (\mathbf{SV_1} + \mathbf{SV_2})
\end{split}
\end{equation}
\subsubsection{Extracting the set of power injection \textbf{PI} and then inputting as the set of demands for the unbalanced load flow program}
The set of power injection includes loading profile of the pseudo steady-state. Running the unbalanced load flow with such input will produce the results of pseudo steady-state measurements $z_n = \{\mathbf{PI_n}, \mathbf{PF_n}\}$, and pseudo steady-state state variables $x_n = \{\mathbf{SV_n}\}$.
\subsubsection{Conducting element-by-element comparisons then drawing conclusion}
Each corresponding element of the two couple sets $\{\hat{z}, z_n\}$, $\{\hat{x}, x_n\}$ will be compared. Based on our observations from the experimental result for the case of 1-phase equivalent transmission system, the maximum mismatches are always smaller than 1\%. If ALL the comparisons produce the results that are less than this threshold, it is reasonable to conclude that the $\hat{z}$ set will definitely bypass the BDD of SE module.

\section{Case study}
\label{case}
\begin{figure}[h]
	\centering
	\includegraphics[scale=0.5]{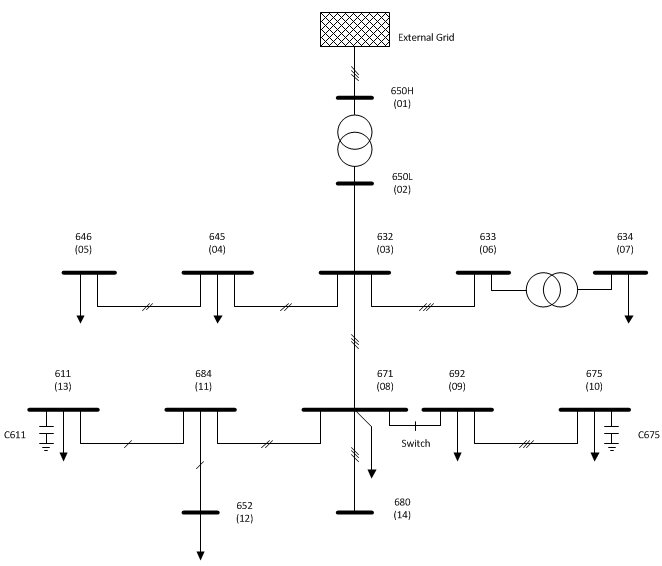}
	\caption{The IEEE 13-node Test Feeder.}
	\label{fig1}
\end{figure}

The IEEE 13-node Test Feeder (Fig. \ref{fig1}) is chosen as the test system following the recommendation from the Test Feeder Working Group of the IEEE’s Distribution System Analysis Subcommittee \cite{schneider2017analytic} for a task related to state estimation. It has eleven overhead lines and underground cables with seven different configurations for various 1-, 2-, or 3-phase laterals.  This system load is diverse and unbalanced. In this investigation, for the sake of simplicity at early stage of work, the distributed load along the line from node 632 to node 671 is disabled.  The nominal voltage of this distribution network is 4.16 kV line-to-line or 2400 V line-to-neutral. All the mesurement readings and pre-attack state variables are collected from the unbalanced load flow results with the default load profile.

In this section, a case study of the IEEE 13-node Test Feeder is investigated and presented following the order of the proposed attack design scheme. For convenience in representing various quantities, all the nodes will be assigned to new aliases as in Table \ref{table1} (these aliases are also indicated in Fig. \ref{fig1}).
\begin{table}[h!]
	\centering
	\renewcommand{\arraystretch}{1.5}
	\begin{tabular}{|l|c|l|c|}
		\hline 
		\textbf{Node} & \textbf{New Name}    & \textbf{Node} & \textbf{New Name}    \\ [0.3ex] \hline \hline
		650H          & \textit{\textbf{01}} & 650L          & \textit{\textbf{02}} \\ \hline
		632           & \textit{\textbf{03}} & 645           & \textit{\textbf{04}} \\ \hline
		646           & \textit{\textbf{05}} & 633           & \textit{\textbf{06}} \\ \hline
		634           & \textit{\textbf{07}} & 671           & \textit{\textbf{08}} \\ \hline
		692           & \textit{\textbf{09}} & 675           & \textit{\textbf{10}} \\ \hline
		684           & \textit{\textbf{11}} & 652           & \textit{\textbf{12}} \\ \hline
		611           & \textit{\textbf{13}} & 680           & \textit{\textbf{14}} \\ \hline
	\end{tabular}
	\caption{Aliases of nodes in the IEEE 13-node Test Feeder.}
	\label{table1}
\end{table}
\subsection{Attack area}
Adopting the Algorithm \ref{Algo1}, several feasible attack areas are found from the IEEE 13-node system:
\begin{itemize}
	\item $\Omega_{A1}$ = $\{13c, 12a, 11ac, 08ac$, launching on either 13\textit{c} or 12\textit{a}\}. It is categorized as ($a_2$)-type attack area.
	\item $\Omega_{A2}$ = $\{04bc, 05bc$, launching on 05\}. It is categorized as ($a_1$)-type attack area.
	\item $\Omega_{A3}$ = $\{10abc, 09ac, 08b$, launching on 10\}. It is categorized as ($a_1$)-type attack area.
	\item $\Omega_{A4}$ = $\{01bc, 02bc, 03bc, 04bc, 05bc, 06bc, 08bc$, launching on 04\}. It is categorized as ($a_2$)-type attack area.
\end{itemize}
\begin{figure*}[h]
	\centering
	\includegraphics[scale=0.9]{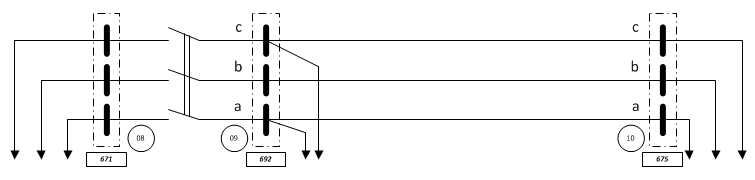}
	\caption{The full 3-phase diagram of attack area $\Omega_{A3}$.}
	\label{fig3}
\end{figure*}
When it comes to the size of attack area, it is apparently that launching an FDI attack on $\Omega_{A4}$ requires much substantial resource than a similar action for $\Omega_{A1}$. Therefore, $\Omega_{A1}$ will be chosen to illustrate the
next steps of the design scheme for ($a_2$)-type attack area. For the case of ($a_1$)-type attack area, $\Omega_{A3}$ almost resembles $\Omega_{A2}$ except it has more phases and is a bit larger since the lack of power injection at node 09\textit{b} gives rise to an expansion towards node 08\textit{b}. In fact, the branch 08-09 is merely a switch, so all the corresponding state variables are identical (as shown in Fig. \ref{fig3}). Thus, working on attack area $\Omega_{A3}$ is considered as a more general case for investigation.

\subsection{Constructing the constraint-based FDI attack model}
\subsubsection{($a_1$)-type Attack Area $\Omega_{A3}$} 
In the traditional distribution network where radial structure dominates, the attack area type of ($a_1$) is omnipresent. The FDI attack here is launched by arbitrarily changing one or several state variables at the initial node while keeping all the other nodes around to be unchanged. The number of changeable state variables at one node is up to six (three voltage magnitudes and three voltage angles for three phases). For the case of attack area $\Omega_{A3}$, we can choose from the set \{$V^a_{10}$, $V^b_{10}$, $V^c_{10}$, $\theta ^a_{10}$, $\theta ^b_{10}$, $\theta ^c_{10}$\}. Meanwhile, the set of state variable at nodes 08 and 09 must be kept unchanged in order to avoid the expansion of attack region.

After imposing one or several initial adjustments onto the selected changeable state variables, the design process goes straight towards the stage of computing malicious measurements. One thing to bear in mind is that no matter how many state variables at node 10 we change, the following power flows must be calculated and then updated to the corresponding meters: {$P^{a,b,c}_{1009}$, $Q^{a,b,c}_{1009}$, $P^{a,b,c}_{0910}$, $Q^{a,b,c}_{0910}$}. After obtaining those malicious values, the power injections at nodes can be calculated. As the load at node 09 has two phases \textit{a} and \textit{c} only, the changes on the line of phase \textit{b} must be justified at node 08 instead.
\subsubsection{($a_2$)-type Attack Area $\Omega_{A1}$}
The full 3-phase diagram of an ($a_2$)-type attack area $\Omega_{A1}$ is illustrated in Fig. \ref{fig5}. In order to keep all the changes happened only inside the attack area, all the state variables at node 08 must be kept unchanged. Consequently, there are 8 changeable state variables within the attack area $\Omega_{A1}$:
\begin{figure*}[h]
	\centering
	\includegraphics[scale=0.75]{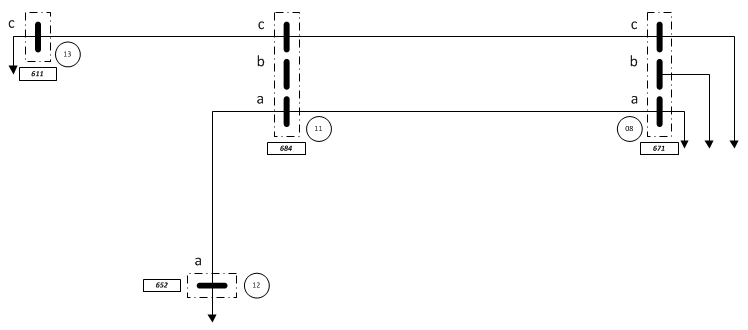}
	\caption{The full 3-phase diagram of attack area $\Omega_{A1}$.}
	\label{fig5}
\end{figure*}
\begin{equation}
\mathbf{SV} = \{V^a_{11}, V^c_{11}, V^a_{12}, V^c_{13}, \theta ^a_{11}, \theta ^c_{11}, \theta ^a_{12}, \theta ^c_{13}\}
\end{equation}
Since the attack area $\Omega_{A1}$ possess only one no-injection node, 11, and it has only two phase (\textit{a} and \textit{c}), the number of no-injection constraint equation related to this node is 4.  
\begin{itemize}
	\item The sum of active power flows of phase \textit{a}:
	\begin{equation}
	\label{eq1}
	\sum P^a_{11} = 0 \Leftrightarrow P^a_{1108n} + P^a_{1112n} = 0
	\end{equation}
	\item The sum of reactive power flows of phase \textit{a}:
	\begin{equation}
	\label{eq2}
	\sum Q^a_{11} = 0 \Leftrightarrow Q^a_{1108n} + Q^a_{1112n} = 0
	\end{equation} 
	\item The sum of active power flows of phase \textit{c}:
	\begin{equation}
	\label{eq3}
	\sum P^c_{11} = 0 \Leftrightarrow P^c_{1108n} + P^c_{1113n} = 0
	\end{equation}
	\item The sum of reactive power flows of phase \textit{c}:
	\begin{equation}
	\label{eq4}
	\sum Q^c_{11} = 0 \Leftrightarrow Q^c_{1108n} + Q^c_{1113n} = 0
	\end{equation}
\end{itemize}
Next, we need to identify the constraint equations related to changes in power injection at nodes and changes in power loss on branches. For the current attack area $\Omega_{A1}$, four constraint equations of this type will be formed:
\begin{itemize}
	\item The sum of all changes in active-related quantities for phase \textit{a} must equal to 0 :
	\begin{equation}
	\label{eq5}
	\begin{split}
	&(P^a_{12n} - P^a_{12o}) + (P^a_{08n} - P^a_{08o}) \\
	&+ (P^a_{L0811n} - P^a_{L0811o}) + (P^a_{L1112n} - P^a_{L1112o})= 0
	\end{split}
	\end{equation}
	\item The sum of changes in reactive-related quantities for phase \textit{a} must equal to 0 :
	\begin{equation}
	\label{eq6}
	\begin{split}
	&(Q^a_{12n} - Q^a_{12o}) + (Q^a_{08n} - Q^a_{08o}) \\
	&+ (Q^a_{L0811n} - Q^a_{L0811o}) + (Q^a_{L1112n} - Q^a_{L1112o})= 0
	\end{split}
	\end{equation} 
	\item The sum of changes in active-related quantities for phase \textit{c} must equal to 0:
	\begin{equation}
	\label{eq7}
	\begin{split}
	&(P^c_{13n} - P^c_{13o}) + (P^c_{08n} - P^c_{08o}) \\
	&+ (P^c_{L0811n} - P^c_{L0811o}) + (P^c_{L1113n} - P^c_{L1113o})= 0
	\end{split}
	\end{equation}
	\item The sum of changes in reactive-related quantities for phase \textit{c} must equal to 0 :
	\begin{equation}
	\label{eq8}
	\begin{split}
	&(Q^c_{13n} - Q^c_{13o}) + (Q^c_{08n} - Q^c_{08o}) \\
	&+ (Q^c_{L0811n} - Q^c_{L0811o}) + (Q^c_{L1113n} - Q^c_{L1113o})= 0
	\end{split}
	\end{equation}
\end{itemize}

In total, the attack model has 8 constraint equations in companion with 8 changeable state variables. At the first sight, it seems that we would have an overdetermined problem to solve because the proposed design scheme requires to devote one state variable for initialization. However, the detailed analysis of the set of constraint equations has shown a different result. First, we have $P^a_{12} = -P^a_{1211}$.

Next, we will delve deeper into the last four constraint equations in order to reveal their true forms. As observed from the detailed 3-phase diagram in Fig. \ref{fig5}, the circuit of phase \textit{c} including \{08, 11, 13\} and the circuit of phase \textit{a} including \{08, 11, 12\} are identical. In each circuit of phase, the equations relevant to active and reactive power are corresponding. Therefore, we just only need to investigate one and then obtain the similar results for the rest.

\begin{figure}[h]
	\centering
	\includegraphics[scale=0.6]{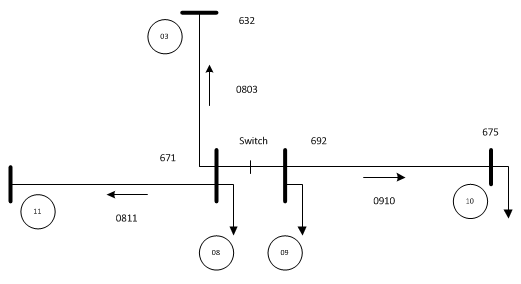}
	\caption{Region around node 08-09.}
	\label{fig6}
\end{figure}

Consider (\ref{eq5}), we already have $P^a_{12n} = -P^a_{1211n}$ and $P^a_{12o} = -P^a_{1211o}$. Next, we focus on the region around node 08-09 (that is extracted and illustrated as in Fig. \ref{fig6}) and obtain the following relationship: $P^a_{08} + P^a_{09} + P^a_{0811} + P^a_{0803} + P^a_{0910} = 0$

As we determine to keep the state variables at 03, 08-09, and 10 unchanged, the two power flows $P^a_{0803}$ and $P^a_{0910}$ will be unchanged accordingly.  Consequently, any change in power injection $P^a_{08}$ will be reflected on $(P^a_{09} + P^a_{0811})$. Let's suppose to keep the metered value of load at node 09, $P^a_{09}$, to be unchanged then all the changes in $P^a_{08}$ will only reflected on $P^a_{0811}$, thus: $P^a_{08n} - P^a_{08o} = -P^a_{0811n} + P^a_{0811o}$. Meanwhile, the branches’ power losses are calculated by adding together two opposite power flows from two terminals. We have $P^a_{L0811n} - P^a_{L0811o} = P^a_{0811n} + P^a_{1108n} - P^a_{0811o} - P^a_{1108o}$ and $P^a_{L1112n} - P^a_{L1112o} = P^a_{1112n} + P^a_{1211n} - P^a_{1112o} - P^a_{1211o}$. Combining all the expressions, then eliminating terms with opposite sign, we have:
\begin{equation}
\label{eq9}
\begin{split}
&(P^a_{1108n} - P^a_{1108o}) + (P^a_{1112n} - P^a_{1112o}) = 0\\
&\Rightarrow(P^a_{1108n} + P^a_{1112n}) - (P^a_{1108o} + P^a_{1112o}) = 0
\end{split}
\end{equation}
We already obtained the value of $P^a_{1108o} + P^a_{1112o}$ from steady state and it exactly equals to 0. Hence, the constraint equation (\ref{eq5}) becomes $P^a_{1108n} + P^a_{1112n} = 0$, or the constraint equation (\ref{eq1}). In conclusion, the requirement for unchanged power in the attack area has already been fulfilled by the requirement for energy conservation at zero-injection node. Also, applying the same procedure as above, the constraint equations (\ref{eq6}), (\ref{eq7}), and (\ref{eq8}) all have reduced forms resemble to the constraint equations (\ref{eq2}), (\ref{eq3}), and (\ref{eq4}). Consequently, there are only four constraint equations (\ref{eq1})-(\ref{eq4}) for this attack area.

At this point, there are \textbf{four} constraint equations and \textbf{eight} changeable state variables. Thus, we can launch an attack by initializing an adjustment on a state variable arbitrarily, keeping three others to be unchanged, and then formulating the four constraint equations with the rest four state variables. The problem of designing an FDI attack now requires solving a set of four nonlinear constraint equations for four unknowns. To illustrate the process, we will choose voltage angle of phase \textit{a} at node 12, $\theta ^a_{12}$, as the initial point to launch an attack. In addition, three state variables will be kept unchanged are $V^a_{11}$, $V^c_{11}$, and $\theta ^c_{11}$. Finally, we have to solve a set of constraint equations \{(\ref{eq1})-(\ref{eq4})\} for four unknowns \{$V^a_{12}$, $V^c_{13}$, $\theta ^a_{11}$, $\theta ^c_{13}$\}. Before generating the set of manipulated state variables, a care must be taken to guarantee the correctness of results. Therefore, we must examine the result of steady state first. It means that we will launch an attack with an initial adjustment amount of zero to the state variable $\theta ^a_{12}$. As the solving process for the set of constraint equations yielded a set of state variables that matches exactly with the steady state values gathered from running load flow program, it is certain that the set of constraint equations is formed appropriately.

\section{Experimental Results}
\label{exre}
The outcomes of the attack model in the previous section are the sets of manipulated state variables which are used to compute malicious measurements. These sets of malicious measurements are then fed into the dedicated Assessment Method (discussed in Section \ref{AMethod}), yielding various experimental results as presented below.
\subsection{Attack area $\Omega_{A3}$}
\begin{figure}[h]
	\centering
	\includegraphics[scale=0.65]{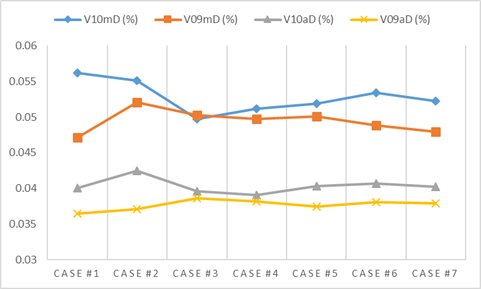}
	\caption{Differences (in percentage) between the state variables obtained from attack design process and from the assessment procedure using PowerFactory 2017.}
	\label{fig4}
\end{figure}

Several experiments are conducted with various small alterations to state variables at node 10. Using the proposed assessment procedure, very small differences between the calculated values and the simulation results are achieved. Fig. \ref{fig4} provides some insight about the average differences (in percentage) between values acquired from attack design scheme and their corresponding results obtained from the assessment procedure. In general, the gaps normally fall in the region from 0.035\% to 0.06\%, which is small enough to negate any dissimilarity. As the loading values from the attack design process can recreate the steady state with a consistent load flow results, the set of malicious measurements is able to gain a "good" reputation from the viewpoint of the SE's BDD module. The detection rate is 0\% since the BDD fails to uncover all the injected malicious measurements. In additionl, because 100\% of these measurements are recognized as normal, the false positive rate is firmly 100\%.
\subsection{Attack area $\Omega_{A1}$}
As the validation process with zero initialization for the constraint-based FDI attack model is accomplished, an attack is now ready to be launched. Extending the work from the Case Study, let's assume the initializing amount on $\theta ^a_{12}$ is +0.1 degree. In practice, the adversaries can prepare for various of different attack cases corresponding to a range of initial values. From such pool of attack cases, a scenario that provides them with the best benefit and economical advantage may be chosen. With the FDI attack model is constructed and solved using MATLAB, we obtain the set of state variables during the attack state as in Table \ref{table2}.
\begin{table}[h]
	\centering
	\begin{tabular}{@{}ccc@{}}
		\toprule
		\textbf{State Var }      & \textbf{Steady State}        & \textbf{Attack State}        \\ \midrule
		$\theta ^a_{12}$ & -5.066 deg          & -4.966 deg          \\
		$V^a_{12}$     & 2.3537 kV/0.9800 pu & 2.3549 kV/0.9805 pu \\
		$V^c_{12}$     & 2.3657 kV/0.9850 pu & 2.3654 kV/0.9848 pu \\
		$\theta ^a_{11}$ & -5.154 deg          & -5.122 deg          \\
		$\theta ^c_{13}$ & 116.517 deg         & 116.518 deg         \\ \bottomrule
	\end{tabular}
	\caption{Comparing state variables from steady state and attack state.}
	\label{table2}
\end{table}

Due to a slight adjustment of +0.1 degree applied for $\theta ^a_{12}$, the values of various state variables are altered accordingly. Although most of changes are minuscule, the gaps between steady state values and attack values are significant, as shown in the second and the third columns of Table \ref{table3}. However, the most important concern is the mismatch between the attack design results and the experimental values obtained from the simulation. Based on the empirical data of SE results for the balanced 1-phase equivalent model under attack, the differences between two individual corresponding member never surpass 1\%. In this case, as indicated by the last column of Table \ref{table3}, the absolute values of difference in percentage are all below 0.3\%. These evidences confirm that the proposed attack design scheme is able to bypass the BDD of the SE module as we achieve the detection rate of 0\% and the false positive rate of 100\% again. In addition, the large deviations between the attacked and the steady state power flows might affect the control actions of the operators, possibly result in incorrect decisions. This issue is discussed in the next section.
\begin{table}[]
	\centering
	\renewcommand{\arraystretch}{1.5}
	\begin{tabular}{|l|r|r|r|r|}
		\hline
		\textbf{Measurement}            & \multicolumn{1}{l|}{\textbf{Steady State}} & \multicolumn{1}{l|}{\textbf{Attack}} & \multicolumn{1}{l|}{\textbf{PF2017}} & \multicolumn{1}{l|}{\textbf{\% Diff}} \\ \hline
		\textit{\textbf{$P^a_{1108}$ (kW)}}   & -123.81                                    & -103.03                              & -103.03                                    & $\sim$0                               \\ \hline
		\textit{\textbf{$P^a_{1112}$ (kW)}}   & 123.81                                     & 103.03                               & 103.03                                     & $\sim$0                               \\ \hline
		\textit{\textbf{$Q^a_{1108}$ (kVAr)}} & -82.84                                     & -105.30                              & -105.23                                    & 0.067                               \\ \hline
		\textit{\textbf{$Q^a_{1112}$ (kVAr)}} & 82.84                                      & 105.30                               & 105.23                                     & 0.067                               \\ \hline
		\textit{\textbf{$P^c_{1108}$ (kW)}}   & -167.83                                    & -172.62                              & -172.61                                    & 0.006                               \\ \hline
		\textit{\textbf{$P^c_{1113}$ (kW)}}   & 167.83                                     & 172.62                               & 172.61                                     & 0.006                               \\ \hline
		\textit{\textbf{$Q^c_{1108}$ (kVAr)}} & 17.83                                      & 11.26                                & 11.29                                      & -0.266                               \\ \hline
		\textit{\textbf{$Q^c_{1113}$ (kVAr)}} & -17.83                                     & -11.26                               & -11.29                                     & -0.266                               \\ \hline
		\textit{\textbf{$P^a_{1211}$ (kW)}}   & -122.93                                    & -102.17                              & -102.17                                    & $\sim$0                               \\ \hline
		\textit{\textbf{$P^a_{0811}$ (kW)}}   & 124.02                                     & 103.25                               & 103.25                                     & $\sim$0                               \\ \hline
		\textit{\textbf{$Q^a_{1211}$ (kVAr)}} & -82.59                                     & -104.99                              & -104.99                                    & $\sim$0                               \\ \hline
		\textit{\textbf{$Q^a_{0811}$ (kVAr)}} & 83.05                                      & 105.49                               & 105.41                                     & 0.076                               \\ \hline
		\textit{\textbf{$P^c_{1311} $ (kW)}}   & -167.45                                    & -172.21                              & -172.21                                    & $\sim$0                               \\ \hline
		\textit{\textbf{$P^c_{0811}$ (kW)}}   & 168.21                                     & 172.99                               & 172.98                                     & 0.006                               \\ \hline
		\textit{\textbf{$Q^c_{1311}$ (kVAr)}} & 18.22                                      & 11.67                                & 11.70                                      & -0.257                               \\ \hline
		\textit{\textbf{$Q^c_{0811}$ (kVAr)}} & -17.58                                     & -10.98                               & -11.01                                     & -0.273                               \\ \hline
		\textit{\textbf{$P^a_{12}$ (kW)}}     & 122.93                                     & 102.17                               & 102.17                                     & $\sim$0                               \\ \hline
		\textit{\textbf{$Q^a_{12}$ (kVAr)}}   & 82.59                                      & 104.99                               & 104.99                                     & $\sim$0                               \\ \hline
		\textit{\textbf{$P^c_{13}$ (kW)}}     & 167.45                                     & 172.21                               & 172.21                                     & $\sim$0                               \\ \hline
		\textit{\textbf{$Q^c_{13} $ (kVAr)}}   & 78.80                                      & 85.32                                & 85.32                                      & $\sim$0                               \\ \hline
		\textit{\textbf{$Q^{Cap}_{13} $ (kVAr)}} & -97.00                                     & -96.99                               & -97.00                                     & -0.01                               \\ \hline
		\textit{\textbf{$P^a_{08}$ (kW)}}     & 382.00                                     & 374.00                               & 374.00                                     & $\sim$0                               \\ \hline
		\textit{\textbf{$Q^a_{08}$ (kVAr)}}   & 204.00                                     & 173.02                               & 173.00                                     & 0.012                               \\ \hline
		\textit{\textbf{$P^c_{08}$ (kW)}}     & 375.00                                     & 366.00                               & 366.00                                     & $\sim$0                               \\ \hline
		\textit{\textbf{$Q^c_{08}$ (kVAr)}}   & 217.00                                     & 229.00                               & 229.00                                     & $\sim$0                               \\ \hline
	\end{tabular}
	\caption{Comparisons of the steady state measurements, the malicious measurements due to an FDI attack, and the simulation results.}
	\label{table3}
\end{table}

While the proposed attack design scheme is able to go through the examination process with zero detection, the majority DC-based design schemes are unable to achieve such perfect stealthy. In the state-of-the-art PowerFactory package, the state estimation module employs nonlinear AC-based model. Hence, the DC-based attack schemes are easily detected as its simplified models cannot consider the nonlinear AC-based characteristics being examined by the PowerFactory’s BDD function. Our experiments have also confirmed this observation. One hundred sets of malicious data to attack the WSCC 9-bus system which is randomly generated by the typical DC-based attack design scheme in \cite{anwar2014vulnerabilities} is examined by the PowerFactory's BDD. In theory, all attack cases can bypass the BDD with no detected bad measurement as these cases have the same normalized residual as of the steady state's measurements: $$\big\|\mathbf{z} - \mathbf{H}\mathbf{x}\big\| = 5.6232\times10^{-4}$$
However, the experiments indicate that all attack cases are ineffective in deceiving the BDD of the PowerFactory. Some cases (54 out of 100) even need to adjust the control parameters for iteration to be converged (scenario attack \#2) \cite{cpe.5956}. As shown in Table \ref{table5}, despite having the same residual, both attack scenario \#1 (converged without adjustment) and \#2 are found to have bad data. It means that while the BDD of PowerFactory cannot detect the existence of bad data generated from the proposed AC-based attack design scheme, it is still robust and effective in dealing with the contemporary DC-based attack schemes.

\begin{table}[]
	\centering
	\renewcommand{\arraystretch}{1.5}
	\begin{tabular}{@{}ccccc@{}}
		\toprule
		\textbf{Scenario}     & \textbf{Quantity} & \textbf{Residual}             & \textbf{BD Detected} & \textbf{Iter. Adjust} \\ \midrule
		\textit{Steady State} & 1                 & $5.6232\times10^{-4}$ & \textbf{Yes}          & No                        \\
		\textit{Attack 1}     & 54                & $5.6232\times10^{-4}$ & \textbf{Yes}          & \textit{\textbf{Yes}}     \\
		\textit{Attack 2}     & 46                & $5.6232\times10^{-4}$ & \textbf{Yes}          & No                        \\ \bottomrule
	\end{tabular}
	\caption{The result of attack an AC-based SE with a DC-based attack scheme.}
	\label{table5}
\end{table}
\section{The impact of the false data injection attack}
\label{imp}
\subsection{Motivations}
\begin{figure*}[h]
	\centering
	\includegraphics[scale=0.68]{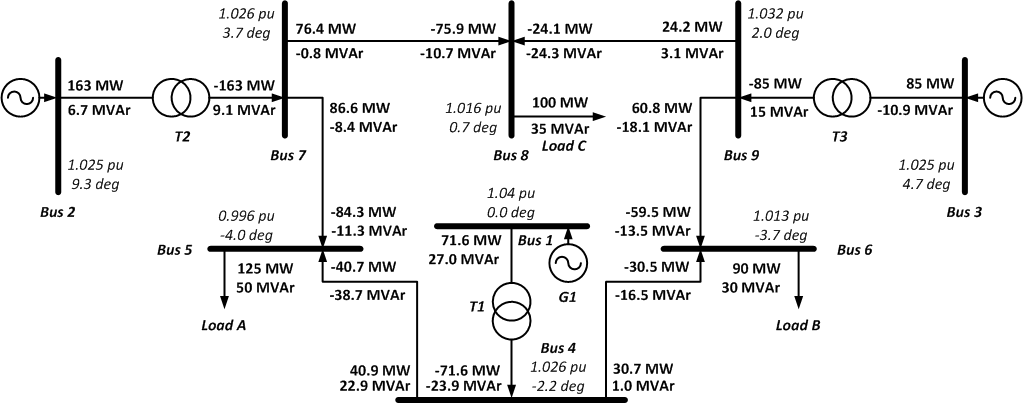}
	\caption{The nine-bus system and its load flow result.}
	\label{fig9}
\end{figure*}
The adversaries have various motivations to organize an FDI attack in distribution grid. Taking advantage of the stealthy capability of the attack, the adversary can disturb the normal operation of the grid (e.g., creating blackout). It only needs to inject the set of false data once, bypassing through the BDD, and then trigger a negative response by the monitoring system to vandalize. In this section, an experiment with the protection system of the WSCC 9-bus system \cite{anderson2008power} is conducted, demonstrating the impact of the FDI attack in disturbing the normal operation of a grid.
\subsection{Impact of FDI attack on the dynamic performance of the 9-bus system}
The WSCC 9-bus system (Fig. \ref{fig9}) \cite{gonzalez2014powerfactory} is used here to conducting dynamic performance experiments as it possesses an adequate set of parameters for dynamic studies. The procedure includes the following steps:
\begin{enumerate}
	\item \textit{Attack design}. Applying the proposed attack design scheme, a set of measurements that contains false data is obtained.
	\item \textit{Data assessment}. The quality of the set of measurements is examined. Only a set that can completely bypass the BDD is selected for the experiment.
	\item \textit{Simulation}. The action of injecting false data with an event of load stepping up (Table \ref{table4}) is simulated. The values of various measurements are adjusted immediately under the effect of this load event.
	\begin{table}[h]
		\centering
		\renewcommand{\arraystretch}{1.5}
		\begin{tabular}{|c|c|c|c|}
			\hline
			\multicolumn{4}{|c|}{\textbf{Variations of loads during the FDI attack}}                    \\ \hline
			\rowcolor[HTML]{68CBD0} 
			\textit{Bus} & \textit{P\_Atk (MW)}   & \textit{P\_SteadyState (MW)}   & \textit{P\_Diff (\%)} \\ \hline
			\textbf{5}   & 240.524856            & 125                           & 94.41989542          \\ \hline
			\textbf{6}   & -69.83155233          & 90                            & -177.5906172         \\ \hline
			\textbf{8}   & 141.8152602           & 100                           & 41.81528127          \\ \hline
			\rowcolor[HTML]{68CBD0} 
			\textit{Bus} & \textit{Q\_Atk (MVAr)} & \textit{Q\_SteadyState (MVAr)} & \textit{Q\_Diff (\%)} \\ \hline
			\textbf{5}   & 17.32483989           & 50                            & -65.35031734         \\ \hline
			\textbf{6}   & 45.21871525           & 30                            & 50.7290606           \\ \hline
			\textbf{8}   & 31.5661642            & 35                            & -9.81095652          \\ \hline
		\end{tabular}
		\caption{Comparison of load values at steady state and attack state.}
		\label{table4}
	\end{table}
	\item \textit{Defining events}. Assuming the current flowing on branch from Bus 4 to Bus 6 (whose the power flows are largest) exceeds the threshold of the overcurrent protection. Circuit breakers at both ends of the branch 4-6 are open.
	\item \textit{Obtaining result}. Observe the responses of the generators' control systems.
\end{enumerate}

\begin{figure*}[]
	\centering
	\includegraphics[scale=0.40]{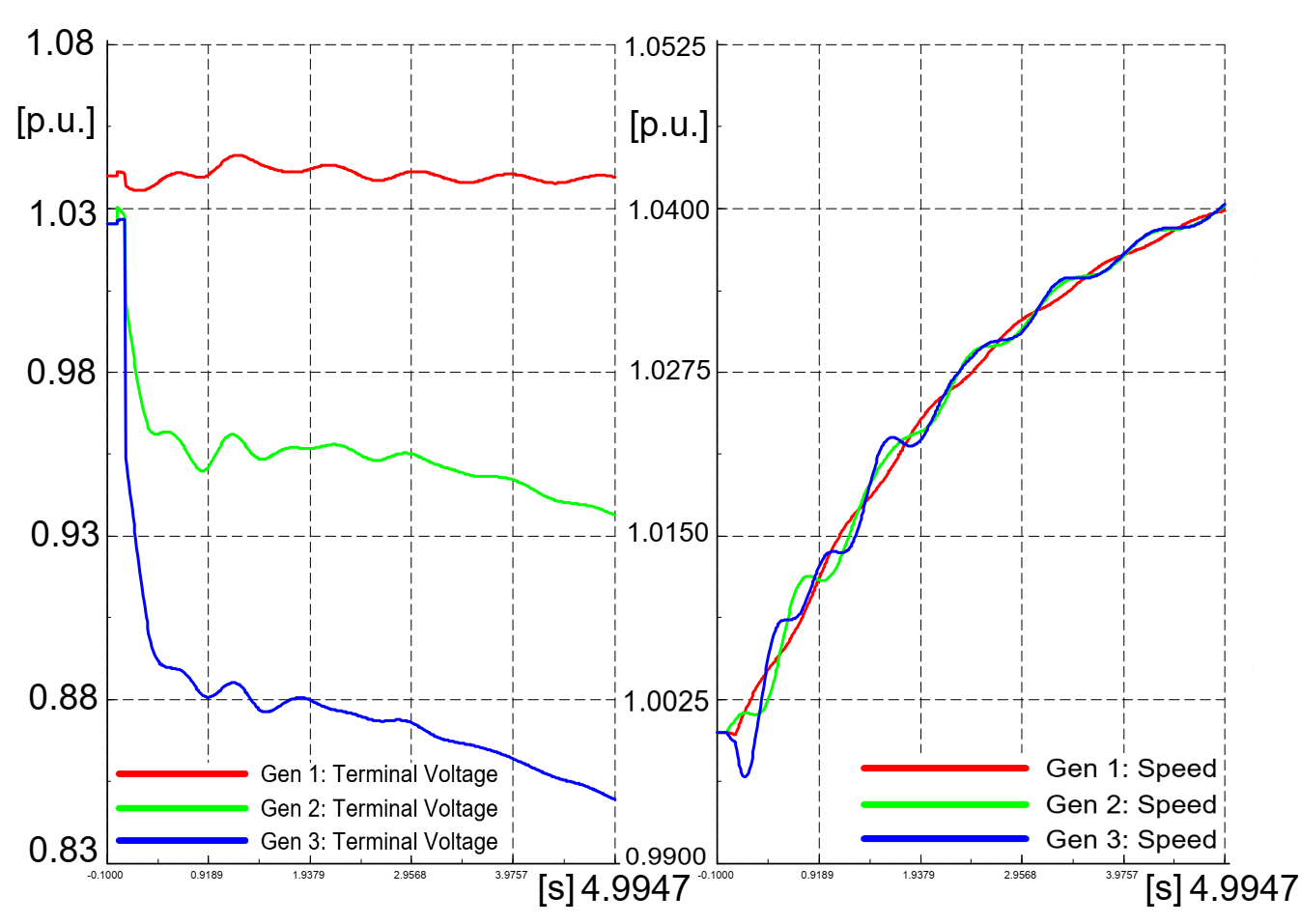}
	\caption{Variations of the generators' speeds and voltages due to injecting false data.}
	\label{fig10}
\end{figure*}

In recent advances in designing system protection, state estimation results are employed directly for protective relays \cite{meliopoulos2016dynamic} \cite{liu2016dynamic} \cite{choi2016effective} \cite{fan2015dynamic} \cite{liu2015dynamic} \cite{albinali2017dynamic}. As the branch 4-6 is tripped the generator speeds gradually increase while the terminal voltages decrease, causing the system to deviate farther from its normal state as shown in Fig. \ref{fig10}. The violation of data integrity in the cyber-domain apparently has influence on operation of the physical environment.
\section{Conclusion}
\label{conclu}
Anticipating the wide deployment of state estimation module in the form of an application of the IoTs edge computing to monitor the emerging smart distribution grid, this paper focuses on the realization of the data-driven cyber-threat false data injection attack.  A comprehensive investigation of constraint-based FDI attack scheme is presented to demonstrate its feasibility and stealthy characteristics. A study case with the IEEE 13-node test feeder illustrates the design process whose results are evaluated by a dedicated procedure. As the set of attack data exhibits the expected ability to bypassing the bad data filtering mechanism, its effectiveness in disrupting the normal operation of power grids is also simulated in another case study with the WSCC 9-bus system.

Although several machine learning based FDI attack detection methods have been proposed \cite{cui2020detecting}, the successful detected FDI attacks are mostly based on simplified settings. To the best of our knowledge, there is no report that such FDI attacks can pass the test of the detection of the industrial commercial tools such as of DIgSILENT's PowerFactory \cite{PF2017}, Nexant's Grid360 \cite{magnago2016multiphase}, Eaton's CYME \cite{krishnan2020validation}, and ETAP \cite{aoufi2020survey}. The success of this proposed attack design scheme under the examination of the professional module has posed a huge challenge to the future operation of smart grid operators. It means that as long as an FDI attack is systematically elaborated (such that it can create a pseudo steady state), all the contemporary data filtering mechanisms will definitely fail to accomplish the assigned missions. Hence, it is a crucial matter to build up a new detection method that has capability to handle this kind of threat. Therefore, one of our future priority work focuses is about a new detection method. In addition, the investigation of attack must be expanded to various alternative scenarios, e.g, when the adversaries have limited attack resources or have no privilege to approach the control center. Such case studies will enhance the knowledge about threats in cyber-physical system, thus greatly contribute to improve the system's security.

\section*{Acknowledgment}
This research is supported by ARC Discovery Grant (IDDP190103660) and ARC Linkage Grant (IDLP180100663).

\bibliographystyle{ieeetran}
\bibliography{B1}
\end{document}